\documentclass[12pt]{iopart}
\usepackage{bm}
\usepackage{epsfig}
\usepackage{graphicx}
\usepackage{color}
\usepackage{iopams}

\newcommand{\edc}{\end{document}}
\newcommand{\bb} {\begin{thebibliography}}
\newcommand{\eb}{\end{thebibliography}}
\newcommand{\be}{\begin{equation}}
\newcommand{\ee}{\end{equation}\normalsize}
\newcommand{\bea}{\begin{eqnarray}}
\newcommand{\eea}{\end{eqnarray}}
\newcommand{\ba}{\begin{array}{l}}
\newcommand{\ea}{\end{array}}
\newcommand{\bc}{\begin{center}}
\newcommand{\ec}{\end{center}}
\newcommand{\dar}{\downarrow}
\newcommand{\upar}{\uparrow}
\newcommand{\re}[1]{(\ref{#1})}
\newcommand{\wh}[1]{\widehat{#1}}
\newcommand{\wt}[1]{\widetilde{#1}}

\begin{document}

\title[Dynamics of a macroscopic spin qubit in spin-orbit coupled BEC]
{Dynamics of a macroscopic spin qubit in spin-orbit coupled Bose-Einstein condensates}

\author{Sh Mardonov$^{1,2,3}$, M Modugno$^{4,5}$ and E Ya Sherman$^{1,4}$}
\address{$^1$ Department of Physical Chemistry, The University of the Basque Country, 48080 Bilbao, Spain}
\address{$^2$ The Samarkand Agriculture Institute, 140103 Samarkand, Uzbekistan}
\address{$^3$ The Samarkand State University, 140104 Samarkand, Uzbekistan}
\address{$^4$ IKERBASQUE Basque Foundation for Science, Bilbao, Spain}
\address{$^5$ Department of  Theoretical Physics and History of Science, University of the Basque Country UPV/EHU, 48080 Bilbao, Spain}

\ead{evgeny.sherman@ehu.eus}
\vspace{10pt}
%\begin{indented}
%\item[] February 2015
%\end{indented}

\begin{abstract}
We consider a macroscopic spin qubit based on spin-orbit coupled Bose-Einstein condensates, where,
in addition to the spin-orbit coupling, spin dynamics strongly depends on the interaction between particles.
The evolution of the spin for freely expanding, trapped, and externally driven
condensates is investigated.  For condensates oscillating at the frequency corresponding to the
Zeeman splitting in the synthetic magnetic field, the spin  Rabi frequency does not
depend on the interaction between the atoms since it produces only internal forces and does not
change the total momentum. However, interactions and spin-orbit coupling bring the system into a mixed
spin  state, where the total spin is inside rather than on the Bloch sphere.
This greatly extends the available spin space making it three-dimensional, but imposes
limitations on the reliable spin manipulation of such a macroscopic qubit.
The spin dynamics can be modified by introducing suitable spin-dependent
initial phases, determined by the spin-orbit coupling, in the spinor wave function.

\end{abstract}


\vspace{2pc}

\submitto{\jpb}
%

\maketitle

\section{Introduction}

The experimental realization of synthetic magnetic fields and spin-orbit coupling (SOC) \cite{EXspielman2009,EXspielman2011}
in Bose-Einstein condensates (BECs) of pseudospin-1/2 particles
has provided novel opportunities for visualizing unconventional phenomena
in quantum condensed matter \cite{EXjin2012,EXji2014}. More recently, also
ultracold Fermi gases with synthetic SOC  have been
produced and studied \cite{EXwangetal2012,EXcheuk2012}.
These achievements have motivated and intense activity, and
a rich variety of new phases and phenomena induced
by the SOC has been discussed both theoretically and experimentally 
\cite{liu2009,galitski2008,stringari2012,zhang2012,zhaih2012,spielman2013,zhang2013,achilleos2013,ozawa2013,wilson2013,lu2013,
xianlong2013,brandon2013,lindong2014}. Recently, it has also been experimentally demonstrated
\cite{EXjin2012,EXqu2013} the ability of a reliable measurement of coupled spin-coordinate motion.

One of the prospective applications of spin-orbit coupled Bose-Einstein condensates
consists in the realization of macroscopic spin qubits \cite{galitski2008}.  A more detailed analysis of
quantum computation based on a two-component BEC was proposed  in \cite{byrnes2012}.
The gates for performing these operations can be produced by means of the SOC and of an external synthetic magnetic field.
Due to the SOC, a periodic mechanical motion of the condensate drives the spin dynamics
and can cause spin-flip transitions at the Rabi frequency depending on the SOC strength.
This technique, known in semiconductor physics as the electric dipole spin resonance,
is well suitable for the manipulation of qubits
based on the spin of a single electron \cite{Nowack,Rashba,You}. For the macroscopic spin qubits based
on Bose-Einstein condensate, the physics is different in at least two aspects. First,
a relative effect of the SOC compared to the kinetic energy can be much stronger
here than in semiconductors. Second, the interaction between the bosons can have a strong
effect on the entire spin dynamics.

Here we study how the spin evolution of a quasi one-dimensional Bose-Einstein condensate depends on the repulsive
interaction between the particles and on the SOC strength. The paper is organized
as follows. In Section 2 we remind the reader the ground state properties of a quasi-one dimensional condensate
and consider simple spin-dipole oscillations. In Section 3 we analyze, by means of the
Gross-Pitaevskii approach,
the dynamics of free,  harmonically trapped, and mechanically driven macroscopic qubits
based on such a condensate. We assume that the periodic mechanical driving resonates with the
Zeeman transition in the synthetic magnetic field and
find different regimes of the spin qubit operation in terms of the interaction between the
atoms, the driving frequency and amplitude. We show that some control of the
spin qubit state can be achieved by introducing phase factors,
dependent on the SOC, in the spinor wave function.
Conclusions will be given in Section 4.

\section{Ground state and spin-dipole oscillations}

\subsection{Ground state energy and wave function}

Before analyzing the spin qubit dynamics, we remind the reader how to obtain the ground state of an interacting BEC.
In particular, we consider a harmonically trapped quasi one-dimensional condensate, tightly bounded
in the transverse directions. The system can be described by the following effective Hamiltonian, where
the interactions between the atoms are taken into account in the Gross-Pitaevskii form:
\begin{equation}
\wh{H}_{0}=\frac{\wh{p}^{2}}{2M}+\frac{M\omega_{0}^{2}}{2}x^{2}+{g}_{1}|\psi(x)|^{2}.
\label{Hho}
\end{equation}
Here $\psi(x)$ is the {condensate} wave function, {$M$ is the particle mass, $\omega_{0}$ is the frequency of the trap 
{(with the corresponding oscillator length $a_{\rm ho}=\sqrt{\hbar/M\omega_0}$)},
and ${g}_{1}=2a_{s}\hbar\omega_{\perp}$ is the effective one-dimensional interaction constant,
with $a_{s}$ being the scattering length of interacting particles, and $\omega_{\perp}\gg\omega_{0}$ being
the transverse confinement frequency.}
For further calculations {we put $\hbar \equiv M\equiv 1$, 
and measure energy in units of $\omega_{0}$ and length in units of $a_{\rm ho}$, respectively.}
All the effects of {the} interaction are determined by the
dimensionless parameter {$\wt{g}_{1}N$, where $\wt{g}_{1}=2\wt{a}_{s}\wt{\omega}_{\perp}$, where $\wt{a}_{s}$ is the 
scattering length in the units of $a_{\rm ho},$ $\wt{\omega}_{\perp}$ is the transverse confinement frequency 
in the units of $\omega_{0}$,} and $N$ is the number of particles. {In physical units, for a condensate of $^{87}{\rm Rb}$
and an axial trapping frequency $\omega_{0}=2\pi\times10$ Hz, the unit of time corresponds to $0.016$ s,
the unit of length $a_{\rm ho}$ corresponds to $3.4\mbox{ }\mu$m,
and the unit of speed $a_{\rm ho}\omega_{0}$ becomes $0.021$ cm/s,
respectively. In addition, considering that $a_{s}=100a_{\rm B}$,
$a_{\rm B}$ being the Bohr radius, in the presence of a transverse confinement with
frequency $\omega_{\perp}=2\pi\times100$ Hz the dimensionless coupling
constant $\wt{g}_{1}$ turns out to be of the order of $10^{-3}$.}

{In order to} find the BEC ground state we minimize the total energy
in a properly truncated harmonic oscillator basis.
To design the wave function we take the real sum of even-order eigenfunctions
\begin{equation}
\psi_{0}(x)=N^{1/2}\sum_{n=0}^{n_{\rm max}}C_{2n}\varphi_{2n}(x).
\label{wf}
\end{equation}
Here
\begin{equation}
\varphi_{2n}(x)=\frac{1}{\sqrt{\pi^{1/2}(2n)!2^{2n}}}
H_{2n}\left({x}\right)\exp\left[-\frac{x^{2}}{2}\right],
\end{equation}
where $H_{2n}\left(x\right)$ are the Hermite polynomials, and the normalization is fixed by requiring that
\begin{equation}
\sum_{n=0}^{n_{\rm max}}C_{2n}^{2}=1.
\end{equation}
The coefficients $C_{2n}$ are determined by minimizing the total energy $E_{\rm tot}$, such that
\begin{equation}
E_{\rm min}=\min_{C_{2n}}\left\{E_{\rm tot}\right\},
\label{Emin}
\end{equation}
where
\begin{equation}
E_{\rm tot}=\frac{1}{2}\int\left[\left(\psi^{\prime}(x)\right)^{2}+x^{2}\psi^{2}(x)+\wt{g}_{1}\psi^{4}(x)\right]\mbox{d}x,
\label{Etot}
\end{equation}
and $|C_{2n_{\rm max}}|\ll 1$.

Formulas \re{Emin} and \re{Etot} yield the ground state energy, while the width of the condensate is defined as:
\begin{equation}
w_{\rm gs}=\left[\frac{2}{N}\int x^{2}|\psi_{0}(x)|^{2}\mbox{d}x\right]^{1/2}.
\label{width}
\end{equation}

\begin{figure}[t]
\bc
\includegraphics[width=.4\textwidth]{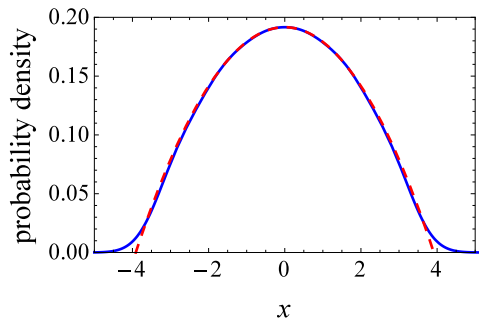}
\caption{{(Color online) {Ground-state probability
density of the condensate obtained from} \re{wf}-\re{Etot} (blue solid line), {compared with} the
Thomas-Fermi approximation in \eref{TF} (red dashed line) for $\wt{g}_{1}N=40.$}}
\label{comparison}
\ec
\end{figure}
In the non interacting limit, $\wt{g}_{1}=0$,  $\psi_{0}(x)$ is the ground state of the harmonic oscillator ($n_{\rm max}=0$), 
that is a Gaussian function with $w_{\rm gs}=1$.
In the opposite, strong coupling limit $\wt{g}_{1}N\gg1$,
the exact wave function \re{wf} is well reproduced (see Figure \ref{comparison}) by the Thomas-Fermi expression
\begin{equation}
\psi_{\rm TF}(x) =\frac{\sqrt{3}}{2}\frac{\sqrt{N}}{w_{\rm TF}^{3/2}}
\left(w_{\rm TF}^{2}-x^{2}\right) ^{1/2};\quad \left|x\right| \le w_{\rm TF},
\label{TF}
\end{equation}
where $w_{\rm TF}=\left(3\wt{g}_{1}N/2\right)^{1/3}.$

In general, for a qualitative description of the ground state one can use instead of the exact wave function \re{wf},
the Gaussian ansatz
\begin{equation}
\psi_{\rm G}(x)=\left(\frac{N}{\pi^{1/2}w}\right)^{1/2}\exp\left[-\frac{x^{2}}{2w^{2}}\right],
\label{gaussian}
\end{equation}
{where the width $w$ is single variational parameter
for the energy minimization.} Then the total energy \re{Etot} becomes:
\begin{equation}
E_{\rm tot}=N\left[\frac{1}{4}\left(w^2+\frac{1}{w^{2}}\right)+
\frac{\wt{g}_{1}N}{2(2\pi)^{1/2}w}\right].
\label{Etotal}
\end{equation}
The latter is minimized with respect to $w$ by solving the equation
\begin{equation}
\frac{dE_{\rm tot}}{dw} = N\left[\frac{1}{2}\left(w-\frac{1}{w^{3}}\right)-
\frac{\wt{g}_{1}N}{2(2\pi)^{1/2}w^{2}} \right] = 0 . %
\label{Eeqq}
\end{equation}
{In the strong coupling regime, $\wt{g}_{1}N\gg 1$, we have $w\gg 1$ so that - to a first approximation - the
kinetic term $\propto 1/w^{3}$ in \re{Eeqq} can be neglected, yielding the following value for the width of the ground state}:
\begin{equation}
\wt{w}_{\rm G} = \left(\frac{\wt{g}_{1}N}{\sqrt{2\pi}}\right)^{1/3}.
\label{w1}
\end{equation}
{The first order correction can be obtained by writing  $w=\wt{w}+\epsilon$ ($\epsilon\ll 1$),
 so that from \re{Eeqq} it follows}:
\begin{equation}
w = \left(\frac{\wt{g}_{1}N}{\sqrt{2\pi}}\right)^{1/3}+\frac{\sqrt{2\pi}}{3\wt{g}_{1}N}.
\label{strongwidth}
\end{equation}
By substituting \re{strongwidth} in \re{Etotal} we obtain {that the
leading term in the ground state energy for $\wt{g}_{1}N\gg1$ is}:
\begin{equation}
E_{\rm min}^{\rm [G]}=\frac{3}{4}N\left(\frac{\wt{g}_{1}N}{\sqrt{2\pi}}\right)^{2/3}.
\label{Eground}
\end{equation}

\begin{figure}[t]
\bc
\includegraphics[width=.4\textwidth]{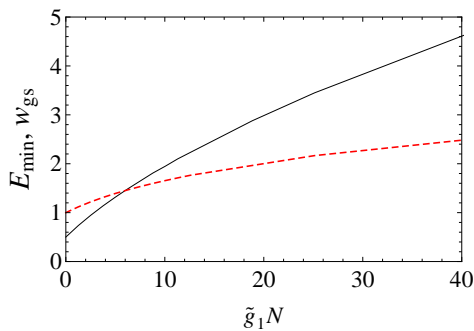}
\caption{(Color online) Ground state energy (black solid line) and {condensate} width (red dashed line) vs. the
interaction parameter $\wt{g}_{1}N$.}
\label{energywidthfreq}
\ec
\end{figure}

In \fref{energywidthfreq} we plot the ground state energy
and the condensate width as a function of the interaction,  as obtained numerically
from \re{Emin} and \re{width}, respectively. As expected, in the strong coupling regime $\wt{g}_{1}N\gg1$ both
quantities nicely follow the behavior (not shown in the Figure) predicted both by the Gaussian approximation and by
the exact solution, namely  $E_{\rm min}\propto (\wt{g}_{1}N)^{2/3}$
and $w_{\rm gs}\propto(\wt{g}_{1}N)^{1/3}$.

\subsection{Simple spin-dipole oscillations}

Let us now turn to the case of a condensate of {pseudospin 1/2 atoms. Here} the system is described by
a two-component spinor wave function $\Psi=\left[\psi_{\uparrow}(x,t),\psi_{\downarrow}(x,t)\right]^T$, still
normalized to the total number of particles $N.$ {The interaction energy (third term in the functional
\re{Etot}) now acquires the form (see, e.g. \cite{stringari2012})}
\begin{equation}
E_{\rm int}=\frac{1}{2}\wt{g}_{1}\int
\left[\left|\psi_{\uparrow}(x,t)\right|^2+\left|\psi_{\downarrow}(x,t)\right|^2\right]^2\mbox{d}x,
\end{equation}
{where, for simplicity and qualitative analysis, we neglect the dependence of interatomic interaction on the spin component ${\uparrow}$
or ${\downarrow}$ and characterize all interactions by a single constant $\wt{g}_{1}.$}

Here we consider spin dipole oscillations, induced by a given small initial symmetric displacement
of the two spin components $\pm\xi$.
For a qualitative understanding, we assume a negligible spin-orbit coupling and a 
Gaussian form of the wave function presented as
\begin{equation}
\Psi_{\rm G}(x)=\frac{1}{\sqrt{2}}
\left[
\ba
\psi_{\rm G}\left(x-\xi\right)\\
\psi_{\rm G}\left(x+\xi\right)
\ea
\right],
\label{gaussianshift}
\end{equation}
where $\psi_{\rm G}$ is given by (\ref{gaussian}), and $\xi\ll w$. {The corresponding} energy is given by:
\begin{equation}
E=E_{\rm min}^{\rm [G]}+E_{\rm sh},
\end{equation}
where $E_{\rm min}^{\rm [G]}$ is defined by \re{Eground} and $E_{\rm sh}$ is the
shift-dependent contribution:
\begin{equation}
E_{\rm sh}=\frac{N}{2}\xi^{2}\left(1-\frac{\wt{g}_{1}N}{\sqrt{2\pi}w^{3}}\right).
\label{energyshift}
\end{equation}
{Then, it follows} that the corresponding oscillation frequency is
\begin{equation}
\omega_{\rm sh}=\sqrt{1-\frac{\wt{g}_{1}N}{\sqrt{2\pi}w^{3}}}.
\label{frequency}
\end{equation}
For strong interaction ($\wt{g}_{1}N\gg 1$) by substituting \re{strongwidth}
in \re{frequency} we obtain:
\begin{equation}
\omega_{\rm sh} \approx \left(\frac{\sqrt{2\pi}}{\wt{g}_{1}N}\right)^{2/3}.
\label{shfrequency}
\end{equation}
Therefore, for strong interaction the frequency of the spin dipole oscillations falls as $(\wt{g}_{1}N)^{-2/3}$,
and {this result is common for the Gaussian ansatz and for the exact solution;
it will be useful in the following section.}

\section{Spin evolution and particles interaction}

\subsection{Hamiltonian, spin density matrix, {and purity}}

To consider the evolution of the driven quasi one dimensional pseudospin-1/2 SOC condensate we begin
with the effective Hamiltonian
\begin{equation}
\wh{H}=\alpha\wh{\sigma}_{z}\wh{p}+\frac{\wh{p}^{2}}{2}+\frac{\Delta}{2}\wh{\sigma}_{x}+
\frac{1}{2}(x-d(t))^2+\wt{g}_{1}\left|\Psi\right|^2.
\label{Htot}
\end{equation}
Here $\alpha$ is the SOC constant {(see \cite{zhaih2012} and \cite{spielman2013} for comprehensive 
review on the SOC in cold atomic gases)}, $\wh{\sigma}_{x}$ and $\wh{\sigma}_{z}$ are the Pauli
matrices, $\Delta$ is the synthetic Zeeman splitting, and $d(t)$ is the driven displacement of
the harmonic trap center as can be obtained by a slow motion of the intersection region of
laser beams trapping the condensate. 

The two-component spinor wave function $\Psi$ is obtained
as a solution of the nonlinear Schr\"{o}dinger equation
\begin{equation}
i\frac{\partial\Psi}{\partial t}=\wh{H}\Psi.
\label{GPE}
\end{equation}
\begin{figure}[t]
\bc
\includegraphics*[width=0.4\textwidth]{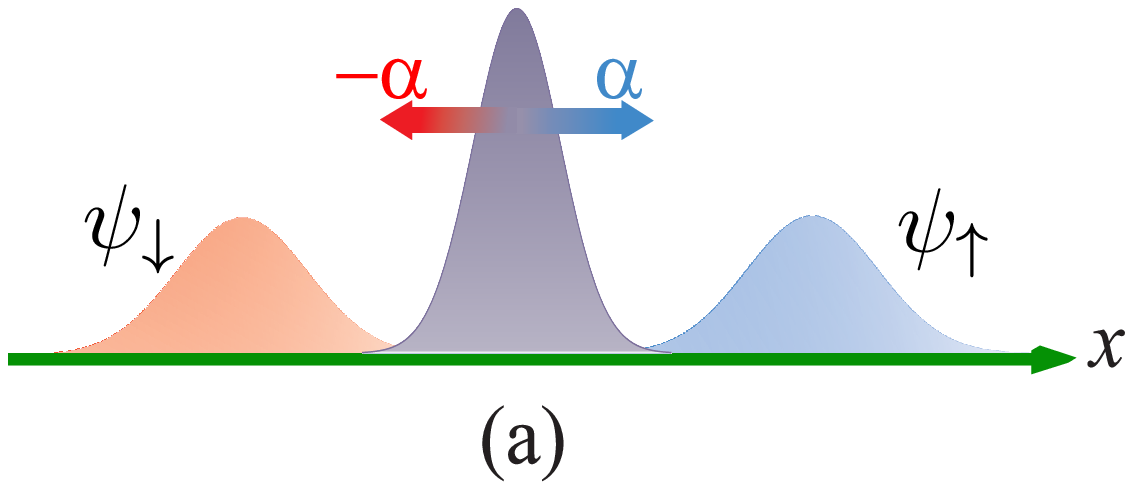}\vspace{1cm}
\includegraphics*[width=0.4\textwidth]{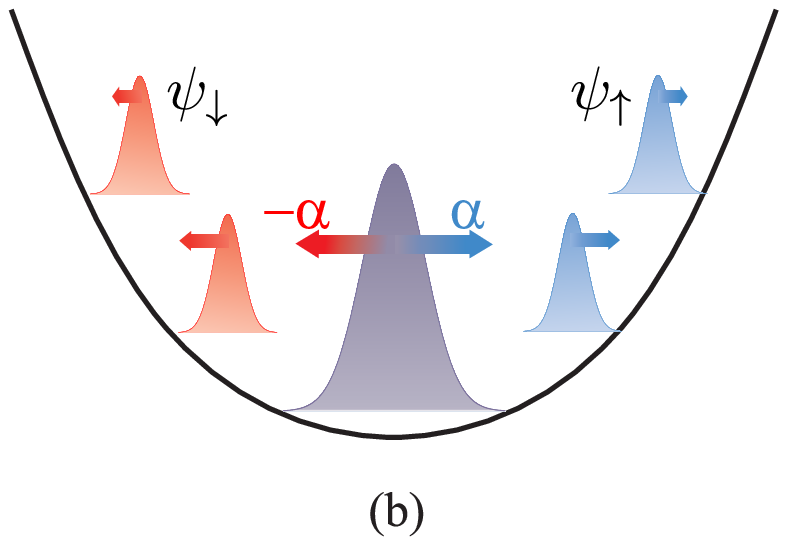}
\caption{(Color online) (a) Separation of {a} freely expanding condensate {in} two spin-up
and spin-down {components} with {opposite} anomalous velocities. (b) Oscillation of the spin-up and spin-down components in the harmonic trap.}
\label{spinoscillation}
\ec
\end{figure}

{To describe spin evolution we introduce the reduced density matrix}
\begin{equation}
{\rho}(t)\equiv |\Psi\rangle\langle\Psi|=
\left[ \ba \rho_{11}(t) \qquad \rho_{12}(t) \\
\rho_{21}(t) \qquad \rho_{22}(t) \ea \right],
\label{densitymatrix}
\end{equation}
where {we trace out the $x-$dependence by calculating integrals} 
\begin{eqnarray}
%\hspace{-.2cm}
&&\rho_{11}(t)=\int|\psi_{\upar}(x,t)|^{2}\mbox{d}x,\ \rho_{22}(t)=\int|\psi_{\dar}(x,t)|^{2}\mbox{d}x,\nonumber\\
&&\rho_{12}(t)=\int\psi_{\upar}^{*}(x,t)\psi_{\dar}(x,t)\mbox{d}x, \quad \rho_{21}(t)=\rho_{12}^{*}(t),
\label{denmat1}
\end{eqnarray}
{and, as a result,} 
\begin{equation}
{\rm tr}\left({\rho}(t)\right)\equiv \rho_{11}(t)+\rho_{22}(t)=N.
\end{equation}
{Although the $|\Psi\rangle$ state is pure, integration in \re{denmat1} produces ${\rho}(t)$
formally describing a mixed state in the spin subspace. One can characterize 
the resulting spin state purity by a parameter $P$ defined as  
\begin{equation}
P=\frac{2}{N^2}\left({\rm tr}\left({\rho}^{2}\right)-\frac{N^2}{2}\right),
\label{purity}
\end{equation}
where $0\le P\le1,$  
\begin{equation}
{\rm tr}\left({\rho}^{2}\right)=N^{2}+2(|\rho_{12}|^{2}-\rho_{11}\rho_{22}),
\label{tracesquare}
\end{equation}
and we omitted the explicit $t-$dependence for brevity.
The system is in the pure state when $P=1,$ that is ${\rm tr}\left({\rho}^{2}\right)=N^2$ with 
$\rho_{11}\rho_{22}=|\rho_{12}|^{2}$.
In the fully mixed state, where ${\rm tr}\left({\rho}^{2}\right)=N^2/2$, we have $P=0$ with
\begin{equation}
\rho_{11}=\rho_{22}=\frac{N}{2}, \qquad \rho_{12}=0.
\label{mixedstate}
\end{equation}
}
The spin components defined by $\langle\wh{\sigma}_{i}\rangle\equiv{{\rm tr}\left(\wh{\sigma}_{i}{\rho}\right)}/{N}$
become
\begin{eqnarray}
&&\langle \wh{\sigma}_{x}\rangle  = \frac{2}{N}{\rm Re}(\rho_{12}),
\quad \langle \wh{\sigma}_{y}\rangle  = -\frac{2}{N}{\rm Im}(\rho_{12}), \nonumber\\
&&\langle \wh{\sigma}_{z} \rangle =\frac{2}{N}\rho_{11}-1,
\label{spincomponent}
\end{eqnarray}
and the purity $P=\sum_{i=1}^{3}\langle \wh{\sigma}_{i} \rangle^{2}$,
{which allows one to match the value of $P$ and the length of the spin vector inside the Bloch sphere}.
For a pure state $\sum_{i=1}^{3}\langle\wh{\sigma}_{i}\rangle^{2}=1$, and
the total spin is on the Bloch sphere. {Instead, for a} fully mixed state
$\sum_{i=1}^{3}\langle\wh{\sigma}_{i}\rangle^{2}=0$, and the spin {null}.
\begin{figure}[t]
\bc
\includegraphics[width=.4\textwidth]{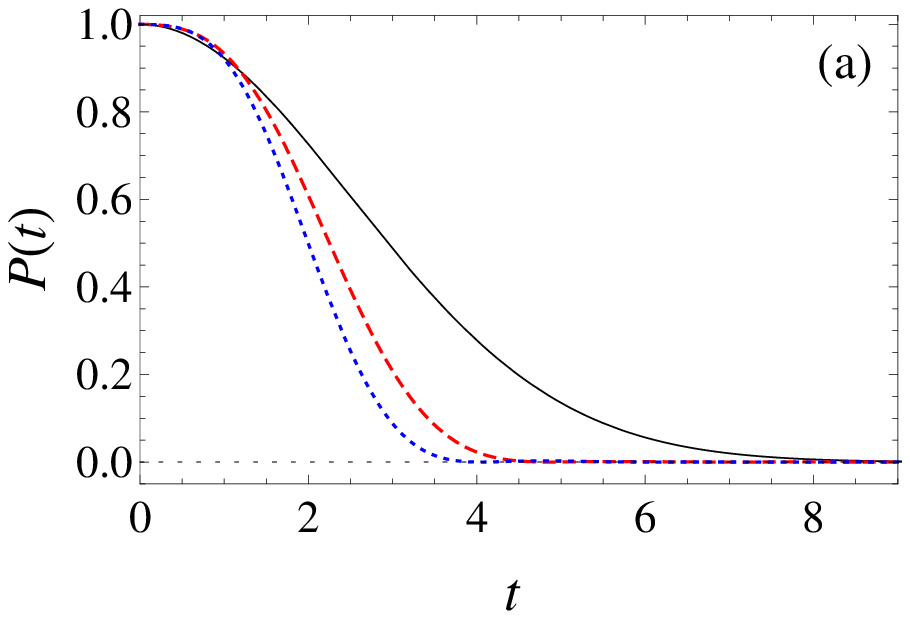}
\includegraphics[width=.41\textwidth]{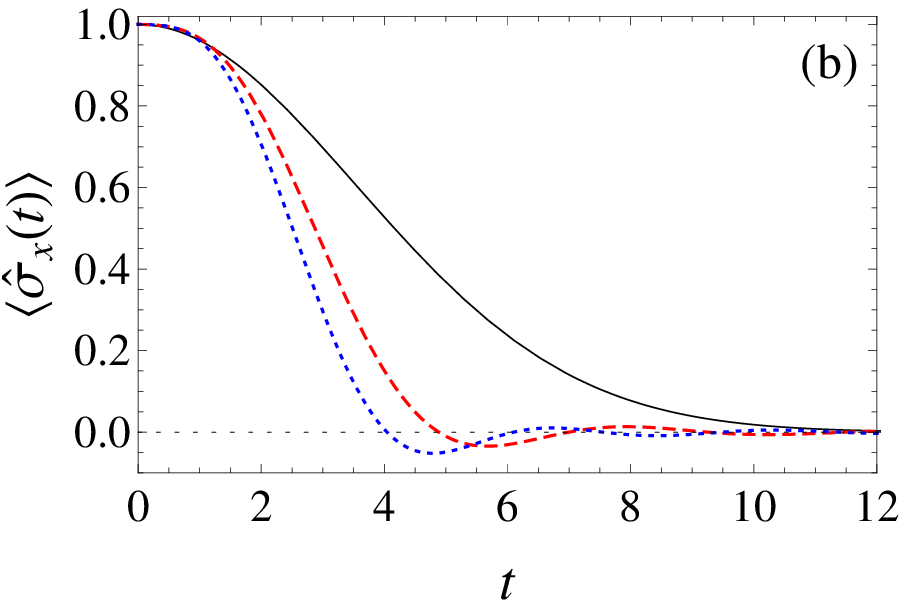}
\caption{(Color online) $(a)$ Purity and $(b)$ spin component
as a function of time for {a} condensate released from {the} trap, for $\alpha=0.2$.
{The lines correspond to $\wt{g}_{1}N = 0$ (black solid line; for the purity cf. \re{purity0}),
$\wt{g}_{1}N = 10$ (red dashed line), and $\wt{g}_{1}N = 20$ (blue dot-dashed line).}
%Black solid line is for $\wt{g}_{1}N = 0$ (see Eq. \re{purity0} for $P_{0}(t)$),  red dashed line is for $\wt{g}_{1}N = 10$,and  blue dot line is for $\wt{g}_{1}N = 20$.
}
\label{trapswitchoff}
\ec
\end{figure}

\subsection{A simple condensate motion}

Let us suppose that a condensate of interacting spin-orbit coupled particles is located
in a harmonic trap and characterized by
an initial wave function
\begin{equation}
\Psi_{0}(x,0)=\frac{1}{\sqrt{2}}\psi_{\rm in}(x)
\left[\ba 1 \\
1
\ea\right],
\label{gswf}
\end{equation}
with the spin parallel to the $x-$axis.

The spin-orbit coupling modifies the commutator corresponding to the velocity operator by introducing the spin-dependent contribution 
as:
\begin{equation}
\wh{v} \equiv i\left[\frac{\wh{p}^{2}}{2}+\alpha\wh{\sigma}_{z}\wh{p},\wh{x}\right] = \wh{p}+\alpha\wh{\sigma}_{z}.
\label{velocityspin}
\end{equation}
The effect of the spin-dependent anomalous velocity term on the condensate motion 
was clearly observed experimentally in \cite{EXjin2012}
as the spin-induced dipole oscillations and in \cite{EXqu2013} as the \textit{Zitterbewegung}.
Since the initial spin in (\ref{gswf}) is parallel to the $x$-axis, the expectation value {of the velocity vanishes,}
$\langle \wh{v} \rangle=0$.

%The following dynamics inside the trap
%is schematically presented in Fig. \ref{spinoscillation}.

Free and oscillating motion of the BEC is shown in \fref{spinoscillation}(a) and \fref{spinoscillation}(b),
respectively. When one switches off the trap, the condensate is set free, 
and the two spin components start to move apart and
the condensate splits up, see \fref{spinoscillation}(a). Each spin-projected component broadens
due to the Heisenberg momentum-coordinate uncertainty and interaction.
The former effect is characterized by a rate proportional to
$1/w_{\rm gs}.$ At large $\wt{g}_{1}N,$ the width $w_{\rm gs}\sim(\wt{g}_{1}N)^{1/3},$
and, as a result, the quantum mechanical broadening rate
decreases as $(\wt{g}_{1}N)^{-1/3}.$ At the same time, the repulsion between the
spin-polarized components accelerates the peak separation \cite{Sherman} and leads to the asymptotic
separation velocity $\sim(\wt{g}_{1}N)^{1/2}$. This acceleration by repulsion leads 
to opposite time-dependent phase factors in $\psi_{\upar}(x,t)$ and $\psi_{\dar}(x,t)$ in (\ref{denmat1}) and, therefore, results 
in decreasing in $\left|\rho_{12}(t)\right|$ and in the purity. Thus, with the increase in the interaction,
the purity and the $x-$ spin component asymptotically
tend to zero faster, as demonstrated in \fref{trapswitchoff}.
For a noninteracting condensate with the initial Gaussian
wave function $\psi_{\rm in}\sim \exp\left(-x^{2}/2w^{2}\right)$ the purity {can be written} analytically as
\begin{equation}
P_{0}(t)=\exp\left[-2\left(\frac{\alpha t}{w}\right)^{2}\right].
\label{purity0}
\end{equation}

\begin{figure}[t!]
\bc
\includegraphics[width=.4\textwidth]{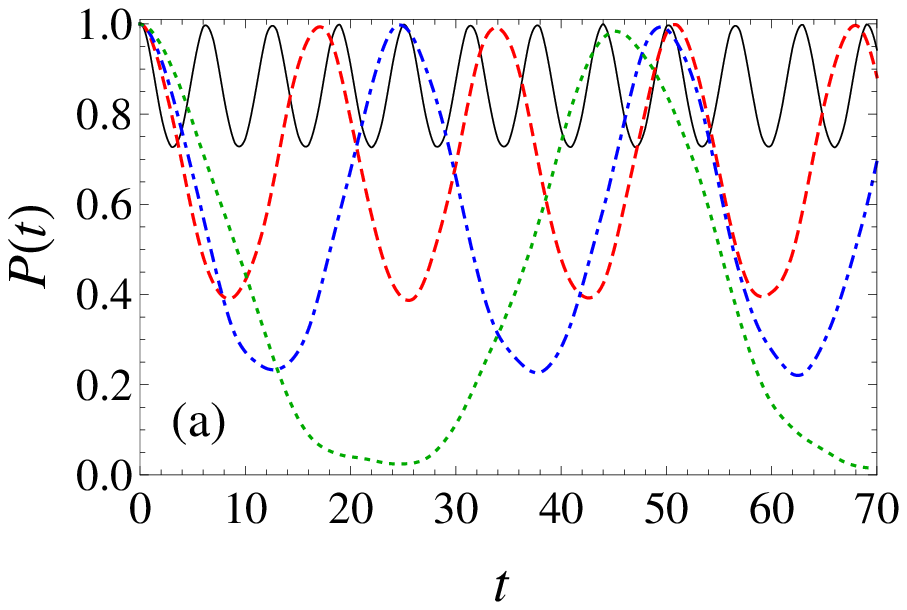}
\includegraphics[width=.4\textwidth]{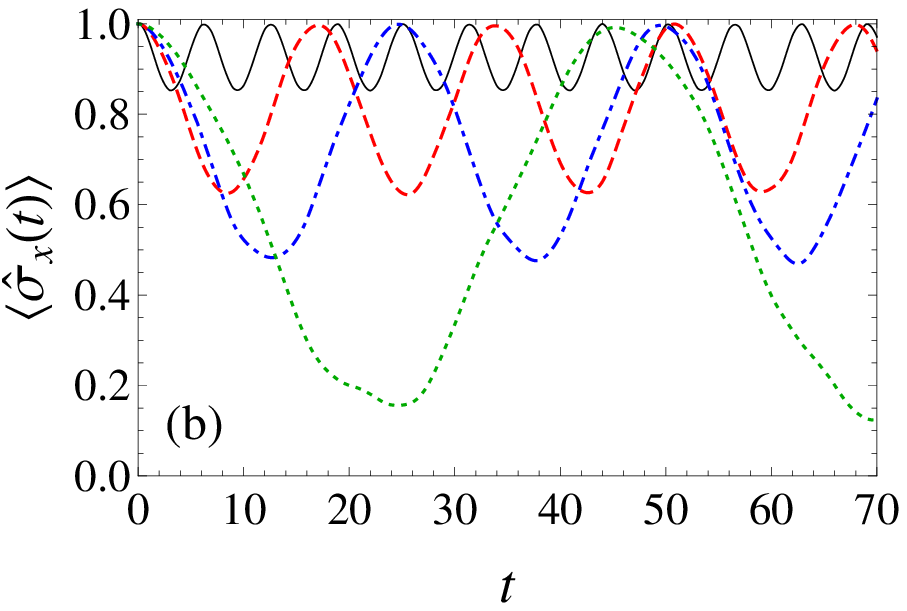}
\includegraphics[width=.4\textwidth]{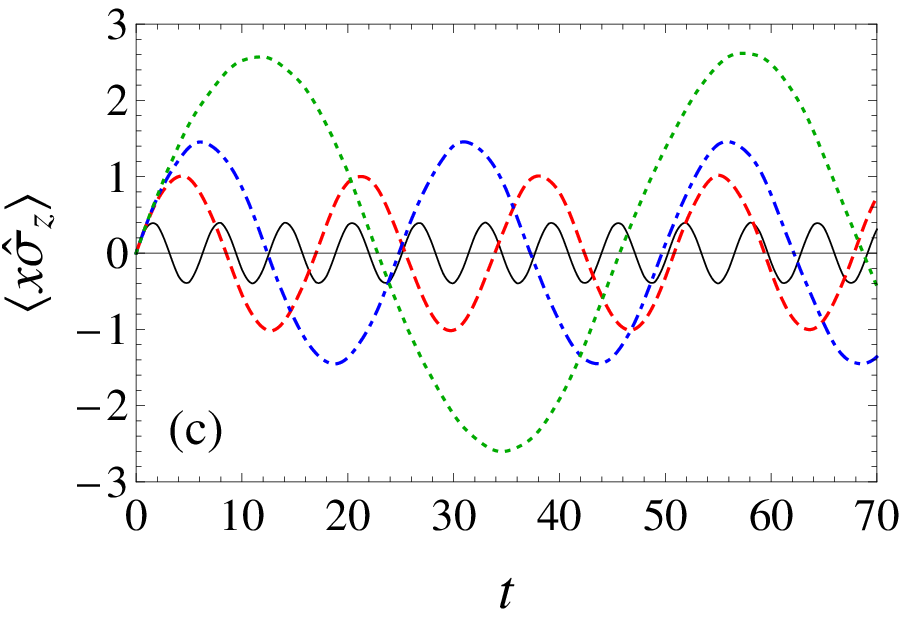}
\caption{(Color online) (a) Purity, (b) spin component,
and (c) spin dipole moment as a function of time for the system in the harmonic trap with
$\alpha=0.2,\ \Delta=0,\ d_{0}=0$.
{The different lines correspond to $\wt{g}_{1}N = 0$ (black solid line),
$\wt{g}_{1}N = 10$ (red dashed line), $\wt{g}_{1}N = 20$ (blue dot-dashed line),
and $\wt{g}_{1}N = 60$ (green dotted line).}
%Black solid line is for $\wt{g}_{1}N = 0$,
%red dashed line is for $\wt{g}_{1}N = 10$, blue dot-dashed line is for
%$\wt{g}_{1}N = 20$, and green dotted line is for $\wt{g}_{1}N = 60$.
}
\label{purityspindipol}
\ec
\end{figure}

In the presence of the trap (\fref{spinoscillation}(b)), the anomalous velocity
in \re{velocityspin} causes spin components (spin-dipole) oscillations
with {a} characteristic frequency of the oder of  $\omega_{\rm sh}$ in \re{shfrequency}.
With the increase in the interatomic interaction, the frequency {$\omega_{\rm sh}$} decreases and, therefore,
the amplitude {of the oscillations arising due to the anomalous velocity ($\sim \alpha/\omega_{\rm sh}$) increases}. 
As a result, {the acceleration and separation of the spin-projected components increase, 
the off-diagonal components of the density matrix in \re{denmat1} become smaller, and} the minimum in $P(t)$ rapidly
decreases to $P(t)\ll1$ as shown by the exact numerical results presented in
figures \ref{purityspindipol}(a) and (b) {\cite{note}}.
In \fref{purityspindipol}(c) we {show} the corresponding evolution of
spin density dipole moment
\begin{equation}
\langle x\wh{\sigma}_{z} \rangle = \frac{1}{N}\int\Psi^{\dag}x\wh{\sigma}_{z}\Psi \mbox{d}x
\label{spintrajectory}.
\end{equation}
Here the oscillation frequency is a factor of two larger than that of the
spin density oscillation.

\subsection{Spin-qubit dynamics and the Rabi frequency}

\begin{figure}[t]
\bc
\includegraphics[width=.4\textwidth]{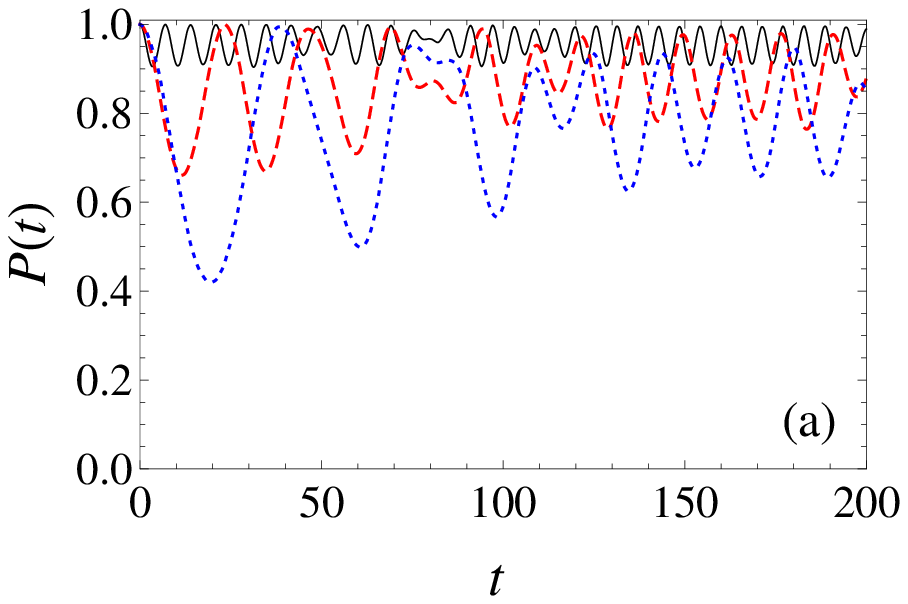}
\includegraphics[width=.4\textwidth]{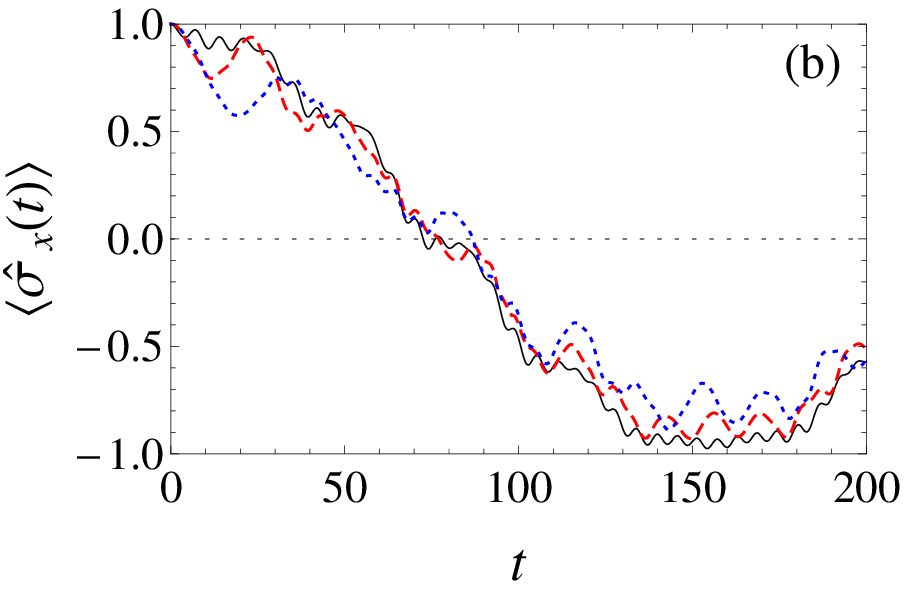}
\caption{(Color online) $(a)$ Purity and $(b)$ spin component
as a function of time for a driven condensate with $\alpha=0.1,\ \Delta=0.1,\ d_{0}=2$.
{The lines correspond to $\wt{g}_{1}N = 0$ (black solid line),
$\wt{g}_{1}N = 10$ (red dashed line), and $\wt{g}_{1}N = 20$ (blue dot-dashed line).}
%Black solid line is for $\wt{g}_{1}N = 0$,  red dashed line is for $\wt{g}_{1}N = 10$,
%and blue dotted line is for $\wt{g}_{1}N = 20$.
}
\label{interaction1}
\ec
\end{figure}

To manipulate the macroscopic spin qubit, the center
of the trap is driven harmonically at the frequency
corresponding to the Zeeman splitting $\Delta$ as
\begin{equation}
d(t)=d_{0}\sin {(t\Delta)},
\label{motiontrap}
\end{equation}
where $d_{0}$ is an arbitrary amplitude and the corresponding spin rotation Rabi frequency $\Omega_{R}$ is defined as 
$\alpha d_{0}\Delta.$ At $\Delta\ll 1,$ as will be considered here, for {a} noninteracting
condensate and a very weak spin-orbit coupling, the
spin component $\langle\wh{\sigma}_{x}(t)\rangle$
{is expected to oscillate approximately as}
\begin{equation}
\langle\wh{\sigma}_{x}(t)\rangle=\cos{(\Omega_{R}t).}
\label{sigmax}
\end{equation}
 The corresponding spin-flip time $T_{\rm sf}$ is
\begin{equation}
T_{\rm sf}=\frac{\pi}{\Omega_{R}}.
\label{frequencyrabi}
\end{equation}

\begin{figure}
\bc
\includegraphics[width=.4\textwidth]{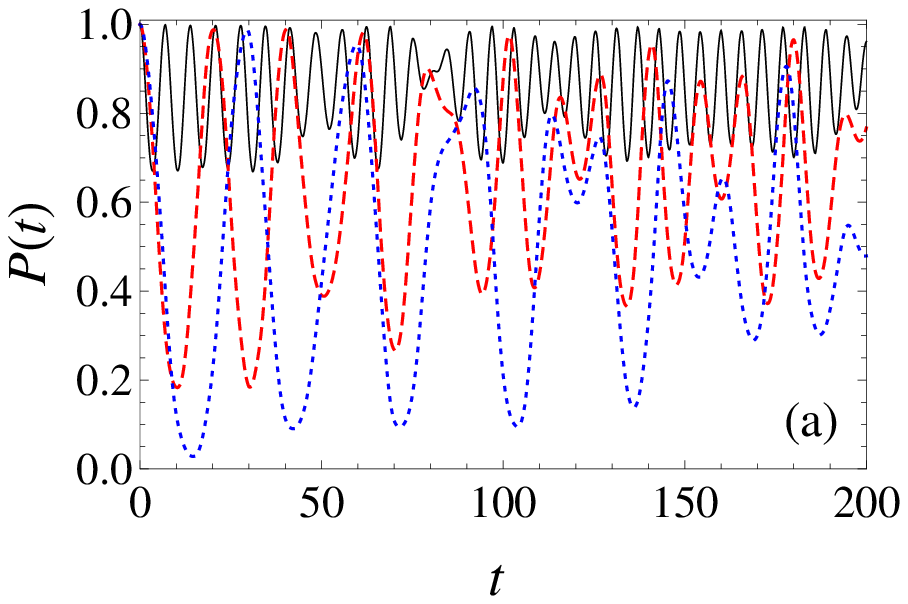}
\includegraphics[width=.4\textwidth]{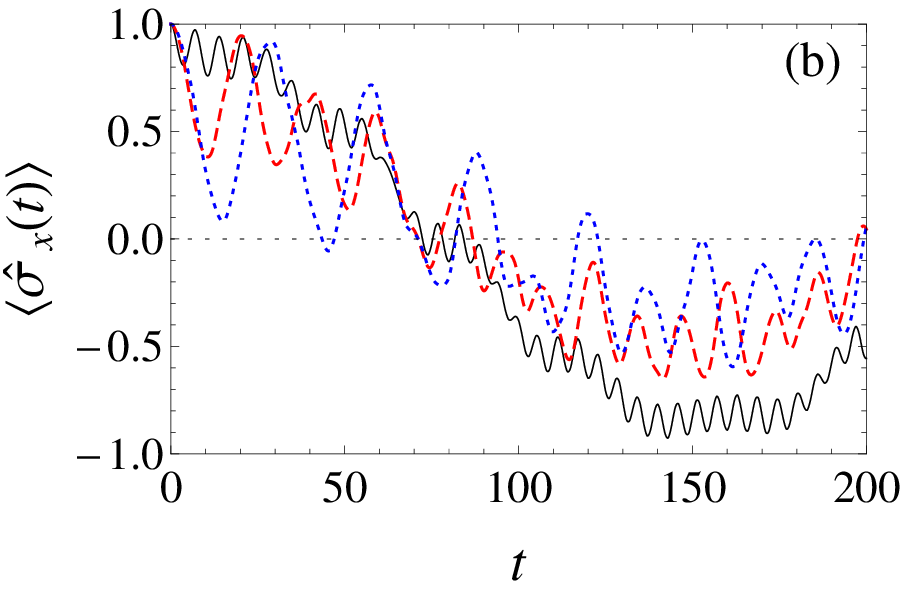}
\includegraphics[width=.4\textwidth]{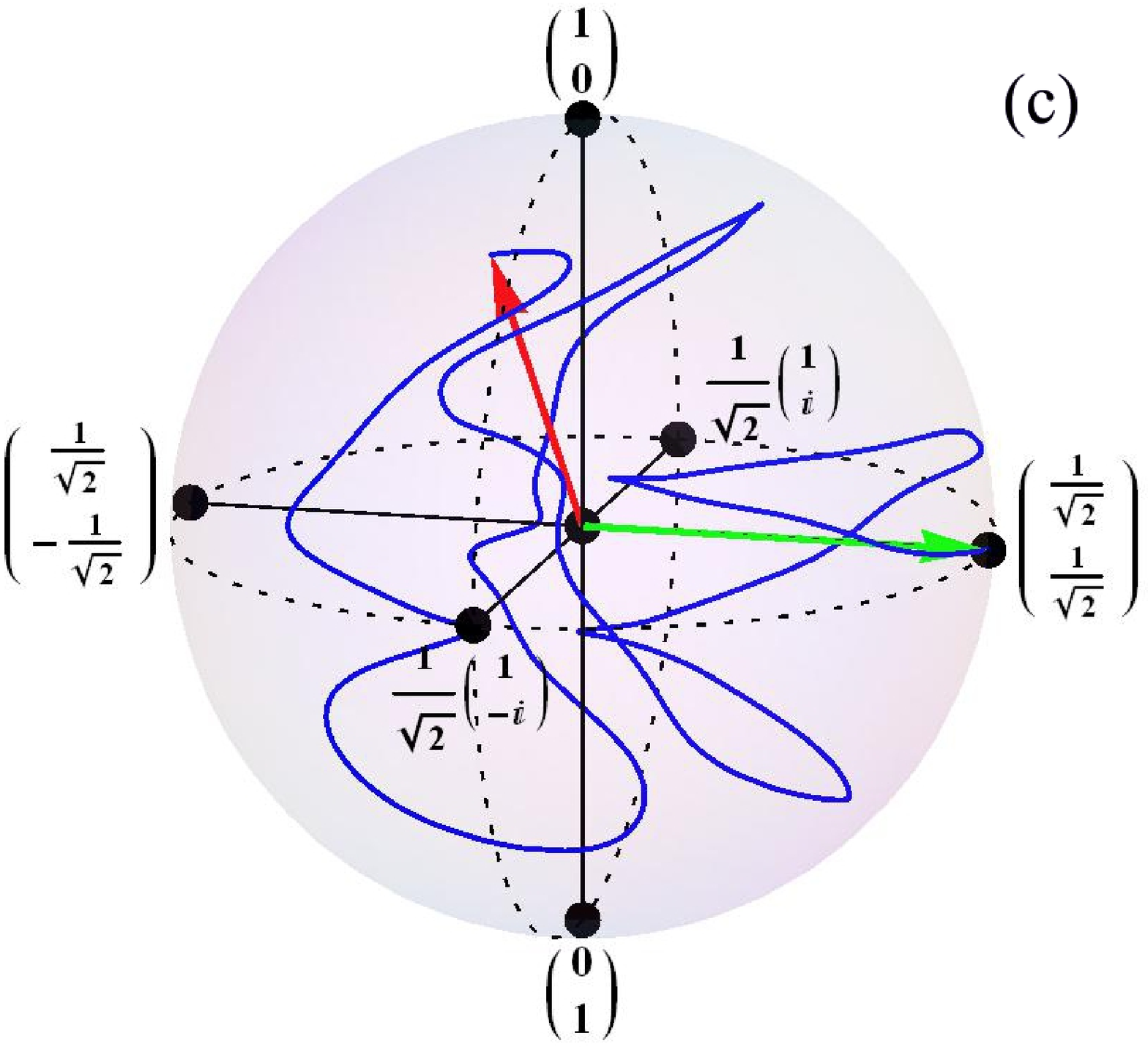}
\caption{(Color online) $(a)$ Purity, $(b)$ spin component,
and $(c)$ spatial trajectory of the spin inside the Bloch sphere
for the driven BEC with $\alpha=0.2,\ \Delta=0.1,\ d_{0}=1$ 
{resulting in the same Rabi frequency as in Figure \re{interaction1}.} In Figures
$(a)$ and $(b)$ the lines correspond to: $\wt{g}_{1}N = 0$ (black solid line),
$\wt{g}_{1}N = 10$ (red dashed line), and $\wt{g}_{1}N = 20$ (blue dotted line).
{At $\wt{g}_{1}N = 0,$ the time dependence
of $\langle\sigma_{x}\rangle$ is rather accurately described by $\cos\left(\Omega_{R}t\right)$ formula,
corresponding to a relatively small variation in the purity, $1-P(t)\ll 1.$  With the increase 
in $\wt{g}_{1}N,$ the purity variation increases and the behavior of $\langle\sigma_{x}\rangle$
deviates stronger from the conventional $\cos\left(\Omega_{R}t\right)$ dependence.}
$(c)$ Here the interaction is {fixed to} $\wt{g}_{1}N = 20$. {The green and red vectors correspond
to the initial and final states of the spin, respectively.
Here the final time is fixed to $t_{\rm fin}=T_{\rm sf}$, see \re{frequencyrabi}.}}
\label{interaction2}
\ec
\end{figure}

Figure \ref{interaction1} shows the time dependence of the purity and the spin
of the condensate for given $\alpha$, $d_{0}$, and $\Delta$ at different interatomic interactions.
In \fref{interaction1}(a) one can see that the increase {of} $\wt{g}_{1}N$
enhances the variation {of} the purity (cf. Fig \ref{purityspindipol}(a)).
This variation prevents a precise manipulation of the
spin-qubit state in the condensate \cite{Khomitsky}.
It follows from \fref{interaction1}(b) that although increasing the interaction
strongly modifies the spin dynamics, it {roughly conserves the spin-flip time
$T_{\rm sf}=50\pi$, see \re{frequencyrabi}}. {To demonstrate the role of the SOC coupling strength $\alpha$ 
and interatomic interaction at nominally the same Rabi frequency $\Omega_{R},$ 
we calculated the spin dynamics presented in Figure \ref{interaction2}.}  By comparing Figures \ref{interaction1}
and \ref{interaction2}(a),(b) we conclude that the increase in the SOC, {at the same Rabi frequency,} 
causes an increase in the variation of the purity and of the spin component.
{These results show that to achieve a required Rabi frequency and a reliable control of the 
spin, it is better to increase the driving amplitude $d_{0}$ rather than the spin-orbit couping $\alpha.$ 
The increase in the SOC strength can result in losing the spin state purity and decreasing the spin length.}
Figure \ref{interaction2}(c) shows the irregular spin evolution of the condensate
inside the Bloch sphere. In \fref{interaction2}(a), for $\alpha=0.2$ and $\wt{g}_{1}N=20,$
the purity decreases almost to zero, placing the spin close
to the center of the Bloch sphere, as can be seen in {\fref{interaction2}(c)}.
It follows from Figures \ref{interaction1}(b) and \ref{interaction2}(b) that {in order}
to protect pure macroscopic spin-qubit states, {the Rabi frequency %\re{frequencyrabi}
should be small. Then,} taking into account that the displacement of the
spin-projected wave packet is of the order of
$\alpha\left(\widetilde{g}_{1}N\right)^{2/3}$ and the packet width is of the order of $\left(\widetilde{g}_{1}N\right)^{1/3}$,
we conclude that for $\alpha\gtrsim\left(\widetilde{g}_{1}N\right)^{-1/3}$, the purity of the driven state tends to zero.
As a result, the Rabi frequency for the pure state evolution {is strongly limited by the interaction between the
particles and} cannot greatly
exceed $d_{0}\Delta/\left(\widetilde{g}_{1}N\right)^{1/3}.$

\begin{figure}[t]
\bc
\includegraphics[width=.4\textwidth]{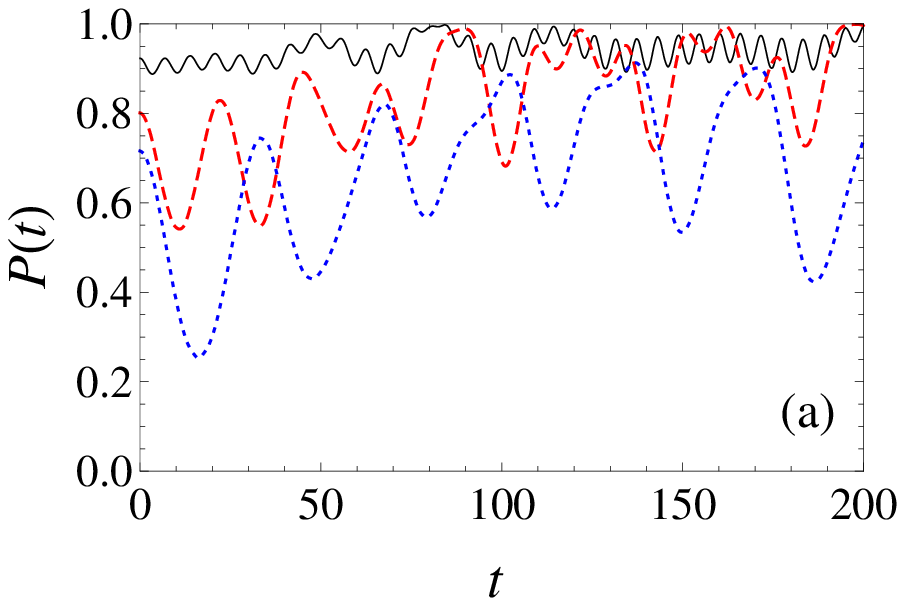}
\includegraphics[width=.4\textwidth]{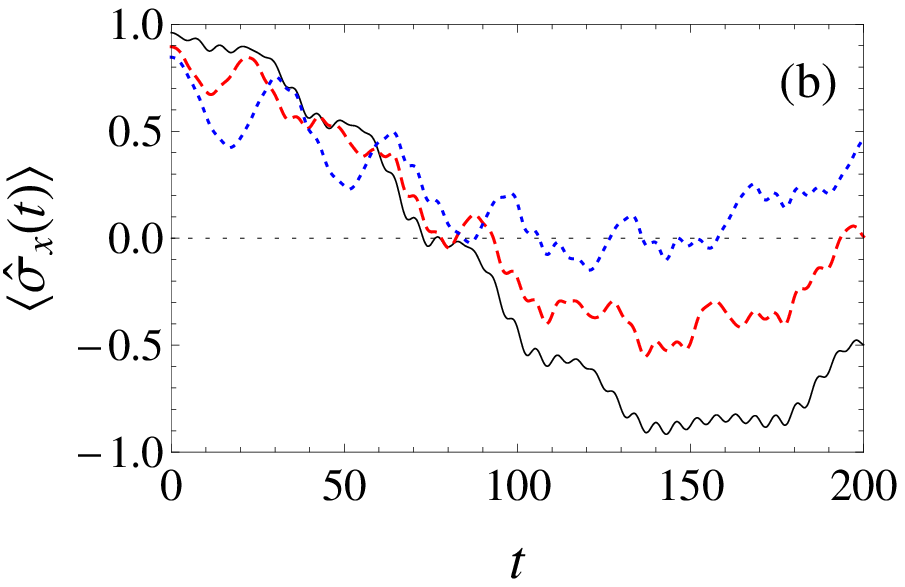}
\includegraphics[width=.4\textwidth]{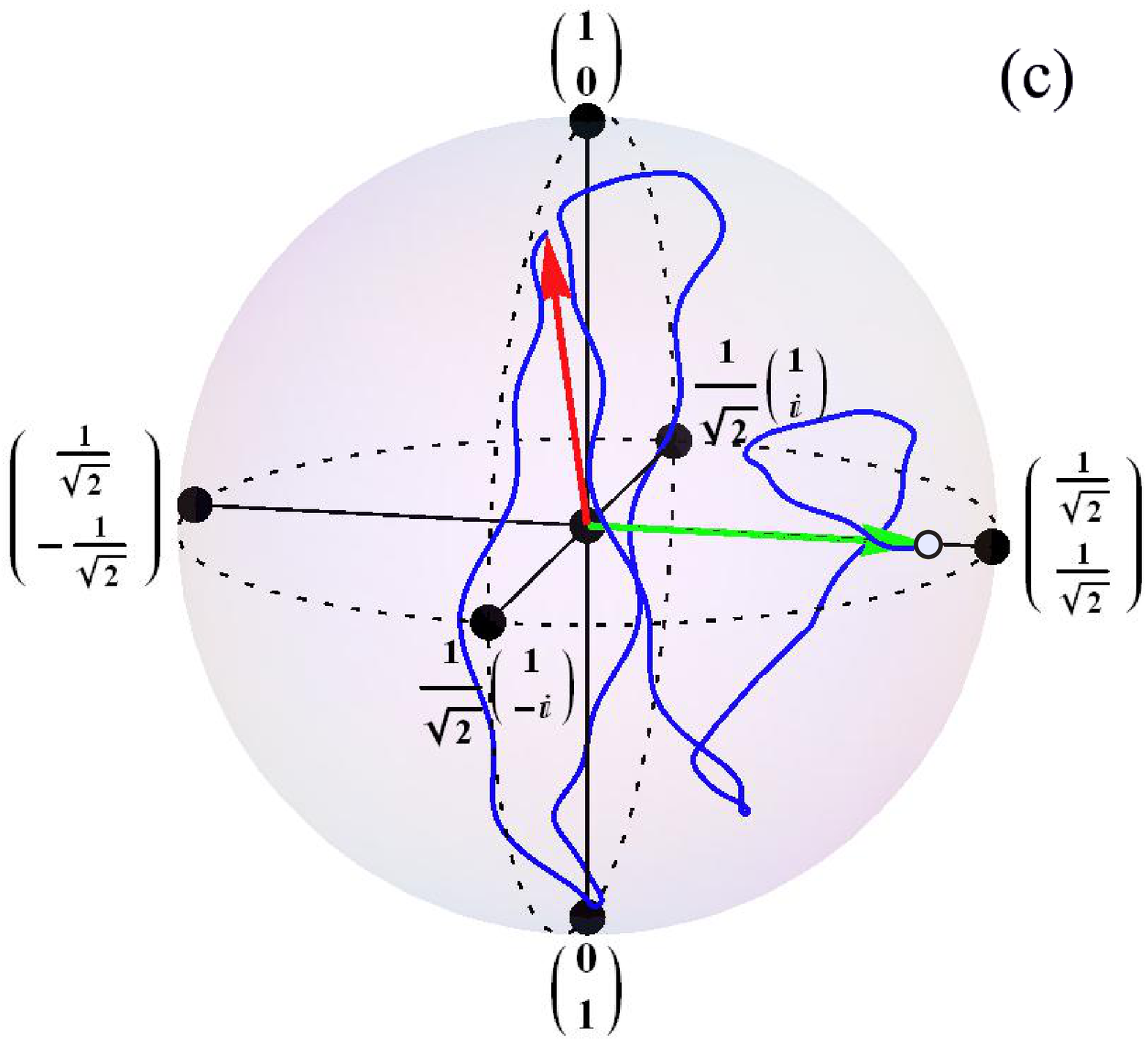}
\caption{(Color online) $(a)$ Purity, $(b)$ spin component,
and $(c)$ trajectory of the spin inside the Bloch sphere for {a} driven BEC
with initial phases as in \re{gswf1} and $\alpha=0.2,\ \Delta=0.1,\ d_{0}=1.$ 
In Figures $(a)$ and $(b)$
the black solid line is for $\wt{g}_{1}N = 0$, {the} red dashed line is for $\wt{g}_{1}N = 10$, 
and {the} blue dotted line is for $\wt{g}_{1}N = 20$. 
In Figure $(c)$ the interaction is $\wt{g}_{1}N = 20$.
The green and red vectors correspond to the initial and final states of the spin, respectively ($t_{\rm fin}=T_{\rm sf}$).
The initial spin state (a solid-line circle with white filling) 
is inside the Bloch sphere since $\langle\sigma_{x}(t=0)\rangle=\sqrt{P(0)}$, and $P(0)<1$ due to the 
mixed character in the spin subspace of the state in \re{gswf1}.} 
\label{phaseinteraction2}
\ec
\end{figure}

{In addition, it is interesting {to note that for} $\widetilde{g}_{1}N\gg 1,$ where the spin
dipole oscillates at the frequency of the order of $\left(\widetilde{g}_{1}N\right)^{-2/3}$ (as given by (\ref{shfrequency})), 
the perturbation due to the trap motion is in the high-frequency limit already at $\Delta\ge\left(\widetilde{g}_{1}N\right)^{-2/3}$, 
having a qualitative influence on the spin dynamics \cite{Bukov,Jorge,Xiong}.}

\subsection{Phase factors due to spin-orbit coupling}

The above results show that the spin-dependent velocity in \re{velocityspin},
{ along with the interatomic repulsion, results in decreasing the spin state purity and}
produces irregular spin motion inside the Bloch sphere. To reduce the effect of
these anomalous velocities and to prevent the {resulting} fast
separation {(with the relative velocity of $2\alpha$)} of the spin components,
we compensate them by introducing coordinate-dependent phases {(similar to the Bragg factors)} 
in the wave function \cite{Josephson}.
To demonstrate the effect of these phase factors, we construct the initial spinor $\Psi_{\alpha}(x,0)$
by a coordinate-dependent SU(2) rotation \cite{Tokatly} of the state
with $\langle\sigma_{x}\rangle=1$  in \re{gswf} as
\begin{equation}
\Psi_{\alpha}(x,0)=e^{-i\alpha x\wh{\sigma}_{z}}\Psi_{0}(x,0)=
\frac{\psi_{\rm in}(x)}{\sqrt{2}}\left[\ba e^{-i\alpha x} \\ e^{i\alpha x}  \ea \right].
\label{gswf1}
\end{equation}
{The expectation value of the velocity \re{velocityspin} at each component $\psi_{\rm in}(x)\exp({\pm i\alpha x})$
is zero, and, as a result, the $\alpha$-induced separation of spin components vanishes, making, as can be easily seen \cite{Tokatly},
the spinor \re{gswf1} the stationary state of the spin-orbit coupled BEC in the Gross-Pitaevskii approximation.} 

{In terms of the spin density matrix \re{denmat1}, the state \re{gswf1} is mixed.}
For a Gaussian condensate with the width $w$,
we get the following expression for the purity at $t=0$
\begin{equation}
P_{\alpha}^{\rm [G]}(0)=\exp\left[-2\left(\alpha w\right)^{2}\right].
\label{newpurity}
\end{equation}
Instead, in the case of a Thomas-Fermi wave function as in \re{TF}, in the limit $\alpha w_{\rm TF}\gg 1$ the initial purity behaves as
\begin{equation}
P_{\alpha}^{\rm [TF]}(0)\sim \frac{\cos ^{2}(2\alpha w_{\rm TF})}{\left(\alpha w_{\rm TF}\right)^{4}}.
\end{equation}
Both cases are characterized by a rapid decrease as $\alpha w_{\rm TF}$ is increased \cite{You}.

{In the absence of external driving, the spin components and purity of \re{gswf1} state remain constant. 
With the driving, spinor components evolve with time and the observables
show evolution quantitatively different from that presented in Figure \ref{interaction2}.}
In figure \ref{phaseinteraction2} we show the analog
of figure \ref{interaction2} for the initial state in \re{gswf1}, with $\psi_{\rm in}(x)=\psi_{0}(x)$ given by \re{wf}-\re{Emin}.
By comparing these Figures one can see that the inclusion of the spin-dependent
phases in \re{gswf1} strongly reduces the oscillations in the $x-$component of the spin,
making the spin trajectory more regular and {decreasing the variations in the purity $P(t)$ compared to the 
initial state without these phase factors. 

A general effect of the interatomic interaction can be seen in both figures \ref{interaction2} and \ref{phaseinteraction2}. 
Namely, for smaller interactions $\wt{g}_{1}N$, the destructive role of the interatomic repulsion on the spin state purity is reduced and 
the spin dynamics becomes more regular. As a result, at smaller $\wt{g}_{1}N$ the spin trajectory is located closer to the Bloch sphere.}

\section{Conclusions}
We have considered the dynamics of freely expanding and harmonically driven macroscopic spin qubits based on
quasi one-dimensional spin-orbit coupled Bose-Einstein condensates in a synthetic Zeeman field. 
The resulting evolution strongly depends in a nontrivial way on the spin-orbit coupling and interaction between the bosons.
On one hand, spin-orbit coupling leads to the driven spin qubit dynamics. On the other hand,
it leads to a spin-dependent anomalous velocity causing spin splitting of the initial wave packet and reducing
the purity of the spin state
by decreasing the off-diagonal components of the spin density matrix. This destructive influence of 
spin-orbit coupling is enhanced by interatomic repulsion. 
The effects of the repulsion can be interpreted in terms of the increase in the spatial width of the condensate and 
the corresponding decrease
in the spin dipole oscillation frequency with the interaction strength. The joint influence of the repulsion and
spin-orbit coupling can spatially separate and modify the spin components stronger than just the spin-orbit coupling
and result in stronger irregularities in the spin dynamics. The spin-flip Rabi frequency remains, 
however, almost unchanged in the presence of the interatomic interactions
since they lead to only internal forces and do not change the condensate momentum. As a result, to preserve the evolution within a high-purity 
spin-qubit state, with the spin being always close to the Bloch sphere, the spin-orbit coupling
should be weak and, due to this weakness, the spin-rotation Rabi frequency should be small and spin rotation should
take a long time. The destructive effect of both the spin-orbit coupling {and interatomic repulsion} 
on the purity of the spin state can be {controllably and considerably reduced, although not completely removed,}
by introducing spin-dependent Bragg-like phase factors in the initial spinor wave function.

\ack

This work was supported by the University of Basque Country UPV/EHU under
program UFI 11/55, Spanish MEC (FIS2012-36673-C03-01 and
FIS2012-36673-C03-03), and Grupos Consolidados UPV/EHU del Gobierno Vasco
(IT-472-10). S.M. acknowledges EU-funded Erasmus Mundus Action 2 eASTANA,
``evroAsian Starter for the Technical Academic Programme'' (Agreement No.
2001-2571/001-001-EMA2).

%\newpage

%\newpage
%\begin{widetext}
%Supplementary material:
%\end{widetext}


\section*{References}

\begin{thebibliography}{99}
%%%%%%%%%%
%1
\bibitem{EXspielman2009}  Lin Y-J, Compton R L, K Jim\'{e}nez-Garcia, Porto J V and Spielman I B
2009 {\it Nature} {\bf 462} 628
%2
\bibitem{EXspielman2011}   Lin Y-J, Jim\'{e}nez-Garc\'{i}a K and Spielman I B
2011 {\it Nature} {\bf 471} 83
%3
\bibitem{EXjin2012}   Zhang J-Y, Ji S-C, Z Chen, Zhang L, Z-D Du, Yan B, G-S Pan, Zhao B, Deng Y-J, Zhai H, Chen S and Pan J-W
2012 {\it Phys. Rev. Lett.} {\bf 109} 115301
%4
\bibitem{EXji2014}        Ji S-C, Zhang J-Y, L Zhang, Du Z-D, Zheng W, Deng Y-J, Zhai H, Chen Sh and Pan J-W
2014 {\it Nature Physics} {\bf 10} 314
%5
\bibitem{EXwangetal2012}  Wang P, Yu Z-Q, Fu Z, Miao J, Huang L, Chai Sh, Zhai H and Zhang J
2012 {\it Phys. Rev. Lett.} {\bf 109} 095301
%6
\bibitem{EXcheuk2012}     Cheuk L W, Sommer A T, Hadzibabic Z, Yefsah T, Bakr W S and Zwierlein M W
2012 {\it Phys. Rev. Lett.} {\bf 109} 095302
%7
\bibitem{liu2009}        Liu X-J, Borunda M F, Liu X and Sinova J
2009 {\it Phys. Rev. Lett.} {\bf 102} 046402
%8
\bibitem{galitski2008}   Stanescu T D, Anderson B and Galitski V
2008 {\it  Phys. Rev.} A {\bf 78} 023616
%9
\bibitem{stringari2012}  Li Y, Martone G I and Stringari S
2012 {\it EPL} {\bf 99} 56008

                        Martone G I, Li Y, Pitaevskii L P and Stringari S
2012 {\it Phys. Rev.} A {\bf 86} 063621
%10
\bibitem{zhang2012}     Zhang Y, Mao L and Zhang Ch
2012 {\it Phys. Rev. Lett.} {\bf 108} 035302
%11
\bibitem{zhaih2012}     Zhai H 2012 {\it Int. J. Mod. Phys.} B {\bf 26} 1230001
%12
\bibitem{spielman2013}  Galitski V and Spielman I B
2013 {\it Nature} {\bf 494} 49
%13
\bibitem{zhang2013}    Zhang Y, Chen G and Zhang Ch
2013 {\it Scientific Reports} {\bf 3} 1937
%14
\bibitem{achilleos2013} Achilleos V, Frantzeskakis D J, Kevrekidis P G and Pelinovsky D E
2013 {\it Phys. Rev. Lett.} {\bf 110} 264101
%15
\bibitem{ozawa2013}     Ozawa T,  Pitaevskii L P and Stringari S
2013 {\it  Phys. Rev.} A {\bf 87} 063610
%16
\bibitem{wilson2013}    Wilson R M, Anderson B M and Clark Ch W
2013 {\it Phys. Rev. Lett.} {\bf 111} 185303
%17
\bibitem{lu2013}         L\"{u} Q-Q and Sheehy D E
2013 {\it  Phys. Rev.} A {\bf 88} 043645
%18
\bibitem{xianlong2013}  Xianlong G
2013 {\it  Phys. Rev.} A {\bf 87} 023628
%19
\bibitem{brandon2013}   Anderson B M, Spielman I B and Juzeliu\~{n}as G
2013 {\it Phys. Rev. Lett.} {\bf 111} 125301
%20
\bibitem{lindong2014}  Dong L, Zhou L, Wu B, Ramachandhran B and Pu H
2014 {\it  Phys. Rev.} A {\bf 89} 011602
%21
%\bibitem{villasenor2014} Villase\~{n}or B, Zamora-Zamora R, Bernal D and Romero-Roch\'{i}n V
%2014 {\it  Phys. Rev.} A {\bf 89} 033611
%22
\bibitem{EXqu2013}   Qu Ch, Hamner Ch, Gong M, Zhang Ch and Engels P
2013 {\it Phys. Rev.} A {\bf 88} 021604
%23
\bibitem{byrnes2012} Byrnes T, Wen K and Yamamoto Y
2012 {\it  Phys. Rev.} A {\bf 85} 040306
%24
\bibitem{Nowack}  Nowack K C, Koppens F H L, Nazarov Yu V and Vandersypen L M K
2007 {\it  Science} {\bf 318} 1430
%25
\bibitem{Rashba} Rashba E I and Efros Al L
2003 {\it Phys. Rev. Lett.} {\bf 91} 126405

\bibitem{You} It is interesting to mention that an increase in spin-orbit coupling strength
after a certain value leads to a less efficient spin driving as shown with perturbation 
theory approach by Li R, You J Q, Sun C P and Nori F
2013 {\it Phys. Rev. Lett.} {\bf 111} 086805

%26
\bibitem{Sherman} This procedure models the von Neumann quantum spin measurement, see 
 Sherman E Ya and Sokolovski D
 2014 {\it New J. Phys.} {\bf 16} 015013
%27
\bibitem{note} An analysis of the spin-dipole oscillations based on the sum rules was presented in \cite{stringari2012}
%28
\bibitem{Khomitsky} This low precision of the spin control can be seen as a general feature of the 
systems where external perturbation strongly drives the orbital motion. See, e.g.
Khomitsky D V, Gulyaev L V and Sherman E Ya 2012 {\it Phys. Rev.} B {\bf 85} 125312

%29
\bibitem{Bukov} A technique to study and engineer the high-frequency behavior has been introduced in
Bukov M, D'Alessio L and Polkovnikov A 2014 arXiv:1407.4803. Its application to the spin dynamics of the spin-orbit coupled 
BEC goes far beyond the scope of this paper, however.

%30
\bibitem{Jorge} Optimal control theory proposed by Budagosky J A and Castro A 2015 {\it The European Physical Journal} B {\bf 88} 15 
can potentially be applied to the 
engineering of a $d(t)-$dependence more complicated than a harmonic oscillation.

%31
\bibitem{Xiong} Xiong B, Zheng J-H and Wang D-W 2014 arXiv:1410.8444 analyzed the condensate driving in terms of the multichannel quantum interference.

%32
\bibitem{Josephson} These phase-related Josephson effect in the spin-orbit coupled BEC in a double-well potential has been recently analyzed in
    Garcia-March M A, Mazzarella G, Dell'Anna L, Juli\'{a}-D\'{i}az B, Salasnich L and Polls A
    2014 {\it  Phys. Rev.} A {\bf 89} 063607

    Citro R and Naddeo A 2014 arXiv:1405.5356

%33
\bibitem{Tokatly} Tokatly I V and Sherman E Ya 2010 {\it Phys. Rev.} B {\bf 82} 161305


%\bibitem{dudarev2004}    A. M. Dudarev, R. B. Diener, I. Carusotto and Q. Niu, {\it Phys. Rev. Lett.} {\bf 92} 153005 (2004).
%\bibitem{osterloh2005}   K. Osterloh, M. Baig, L. Santos, P. Zoller and M. Lewenstein, {\it Phys. Rev. Lett.} {\bf95} 010403 (2005).
%\bibitem{ruseckas2005}   J. Ruseckas, G. Juzeli\={u}nas, P. \"{O}hberg and M. Fleischhauer, {\it Phys. Rev. Lett.} {\bf 95} 010404 (2005).

%
%\bibitem{dalibard2010}   G. Juzeli\={u}nas, J. Ruseckas and J. Dalibard, Phys. Rev. A. {\bf 81} 053403 (2010).
%\bibitem{wang2010}       C. Wang, C. Gao, C-M. Jian and H. Zhai, {\it Phys. Rev. Lett.} {\bf 105} 160403 (2010).
%\bibitem{ho2011}         T-L. Ho and S. Zhang, {\it Phys. Rev. Lett.} {\bf 107} 150403 (2011).
%

%\bibitem{zhangatall2012} J-Y. Zang \emph{et al.} {\it Phys. Rev. Lett.} {\bf 109} 115301 (2012).
%\bibitem{anderson2012}   B. M. Anderson, G. Juzeli\={u}nas, V. M. Galitski and I. B. Spielman, {\it Phys. Rev. Lett.} {\bf 108} 235301 (2012).

%\bibitem{ratschbacher2013}   L. Ratschbacher, C. Sias, L. Carcagni, J. M. Silver, C. Zipkes and M. K\"{o}hl, {\it Phys. Rev. Lett.} {\bf 110} 160402 (2013).
%



%
%\end{harvard}
%\endrefs
%

\end{thebibliography}
\end{document}